%% file: main.tex
\documentclass{llncs}

\usepackage{epsfig}
\usepackage{graphics}
\usepackage{theorem}
\usepackage{amssymb}
\usepackage{pstricks}
\usepackage{color}
\usepackage{amsmath}
\usepackage{wrapfig}
\usepackage{subfigure}

{\theoremstyle{plain}
\theoremheaderfont{\normalfont\bfseries}
\theorembodyfont{\rmfamily} \newtheorem{scenario}{Scenario}}

\newcommand{\loeq}{\leqslant}

\newcommand{\bigsum}[2]{\displaystyle\sum_{#1}^{#2}}
\newcommand{\probkn}{P_{\neg {\it drop}}}

\newcommand{\probsn}{P_{{\it select}}}
\newcommand{\probs}{\frac{s}{c}}

\newcommand{\probnotsn}{P_{\neg {\it select}}}
\newcommand{\probdn}{P_{{\it drop}}}
\newcommand{\probdwsh}{\probdn^{|S_A \cap S_B | = \sasb}}
\newcommand{\probdwk}{\probdn^{|S_A \cap C_B | = k}}
\newcommand{\probok}{P_{|S_A \cap C_B | = k}}
\newcommand{\probosh}{P_{|S_A \cap S_B | = \sasb}}
\newcommand{\probdwshp}[1]{\probdn^{|S_A \cap S_B | = #1}}
\newcommand{\corr}{\gamma}
\newcommand{\sminus}{\backslash}
\newcommand{\sasb}{\widehat{s}}

\newcommand{\un}{\!*\!}
\newcommand{\bun}{*}

\title{\mbox{An Analytical Model of Information Dissemination} for a Gossip-based Protocol}
\author{Rena Bakhshi \and Daniela Gavidia \and Wan Fokkink \and Maarten van Steen}
\institute{Department of Computer Science, Vrije Universiteit Amsterdam, Netherlands\\ \email{\{rbakhshi,daniela,wanf,steen\}@few.vu.nl}}

\begin{document}

\maketitle

\input{abstract}
\input{intro}
\input{protocol}
\input{analysis}
\input{evaluation}
\input{modelling}
\input{conclusion}


\end{document}

%% file: abstract.tex
\begin{abstract}
We develop an analytical model of information dissemination for a gossiping protocol that 
combines both pull and push approaches. With this model we analyse how fast an item is
replicated through a network, and how fast the item spreads in the network, and how fast 
the item covers the network. We also determine the optimal size of the exchange buffer, to 
obtain fast replication. Our results are confirmed by large-scale simulation experiments.
\end{abstract}

%% file: intro.tex
\section{Introduction}
Today, large-scale distributed systems consisting of thousands of nodes are 
commonplace, due to the wide availability of high-performance and low-cost 
devices. Such systems are highly dynamic in the sense that nodes are 
continuously in flux, with new nodes joining and existing 
nodes leaving. 

In practice, large-scale systems are often emulated to discover correlations between design 
parameters and observed behaviour. Such  
experimental results provide essential data on system behaviour. However, they usually show only behaviour of a particular implementation, and can be time consuming. 
Moreover, in general experiments do not give a good understanding of the emergent behaviour of the system, and into how 
parameter settings influence the extra-functional properties of the system. As a result, it is very difficult to predict what the effects of certain design decisions are, as it is practically infeasible to explore the full range 
of input data. A challenge is to develop analytical models that 
capture (part of) the behaviour of a system, and then subsequently optimize 
design parameters following an analytical rather than an experimental approach. 

We are interested in developing and validating analytical models for gossip-based 
systems (cf.\ \cite{1317385}). These systems rely on epidemic techniques for the communication and 
exchange of information. These communication protocols, while having simple 
specifications, show complex and often unexpected behaviour when executed on a 
large scale. Our analytical models of gossip protocols need to be 
realistic, yet, sufficiently abstract to allow for easy prediction 
of systems behaviour. By `realistic' we mean that they can 
be applied to large-scale systems and can capture functional and extra-functional 
behaviour such as replication, coverage, convergence, and other system dynamics (see
\cite{eugster.p2004}). Such models are amenable for mathematical 
analysis, to make precise predictions. Furthermore, we will exploit the fact that 
because an analytical model presents an abstraction of the original protocol, a 
simulation of the model tends to be much more efficient (in computation time and memory 
consumption) than a simulation of an implementation of this protocol. 

In this paper, we develop an analytical model of a shuffle protocol from \cite{GVS06},
which was developed to disseminate data items to a collection of wireless devices, in a 
decentralized fashion. A decentralized solution considerably decreases the 
probability of information loss or unavailability that may occur due to a 
single point of failure, or high latency due to the overload of a node. 
Nodes executing the protocol periodically contact each other, according to 
some probability distribution, and exchange data items.
Concisely, a node initiates a contact with its random neighbour, pulls a 
random subset of items from the contacted node, simultaneously 
pushing its own random subset of items. This push/pull approach has a better
performance than a pure push or pull approach \cite{steen2007.15,796561}.
The amount of information exchanged during each contact between two communicating nodes is 
limited. Replication ensures the availability of the data items even in 
the face of dynamic behaviour, which is characteristic of wireless environments. 
Thus, nodes not only conserve the data collectively stored in the network, but 
also relocate it in a random fashion; nodes will eventually see all data items. 

The central point of our study is a rigorous probabilistic analysis of information 
dissemination in a large-scale network using the aforementioned protocol. The behaviour of the protocol 
is modelled on an abstract level as pairwise node interactions. When two neighbouring 
nodes interact with each other (gossip), they may undergo a state transition (exchange 
items) with a certain probability. The transition probabilities depend on the probability 
that a given item in a node's cache has been replaced by another item after the shuffle.
We calculated accurate values for these probabilities, yielding a rather
complicated expression. We also determined a close approximation that is expressed by a much 
simpler formula, as well as a correction factor for this approximation, allowing for precise 
error estimations. Thus we obtain a better understanding of the emergent behaviour of the
protocol, and how parameter settings influence its extra-functional behaviour.  

We investigated two properties characterizing the protocol, namely, the number of replicas of a given item 
in the network at a certain moment in time (replication), and the number of nodes 
 that have `seen' this item over time (coverage). Using the values of the
transition probabilities, we determined the optimal number of items to exchange 
per gossip, for a fast convergence of coverage and replication. 
Moreover, we determined formulas that capture the dissemination of an item in a fully
connected network.
All our modelling and analysis results are confirmed by large-scale simulations, in which
simulations based on our analytical models are compared with running the actual protocol. 
To the best of our knowledge, we are the first to develop an accurate, realistic formal 
model that can be used to optimally design and fine-tune a given gossip 
protocol. In this sense, our main contribution is demonstrating the feasibility of a 
model-driven approach to developing real-world gossip protocols.

The paper is structured as follows. The remainder of this introduction discusses related
work. Section \ref{sec:protocol} explains the shuffle protocol.
In Section \ref{sec:analysis} the analytical model is developed. Section \ref{sec:expeval}
discusses the results of our experimental evaluations. Section \ref{sec:modelling} presents
a round-based perspective of replication and coverage. And Section \ref{sec:conclusions} contains
the conclusions.

\subsection*{Related work}
Two areas of research are relevant to our paper: rigorous analysis of 
gossip (and related) protocols, and results from mathematical 
theory of epidemics \cite{Bai75,DG99}. The results from epidemics are often used in the 
analysis of gossip protocols \cite{EugGueKerMas04IEEEComp}. We restrict our overview 
to the most relevant publications from the area of gossip protocols.

Several works have focused on gossip-based membership management protocols.

Allavena et al. \cite{1073871} proposed a gossip-based membership
management protocol and analysed the evolution of the number of links 
between two nodes executing the protocol. The states of the associated 
Markov chain are the number of links between pairs of nodes. 
From the designed Markov chain they calculated the expected time 
until a network partition occurs. This case study also includes 
a model of the system under churn. A goal of that paper is to show 
the effect of mixing both pull and push approaches.

Eugster et al.~\cite{eugster.p2003c} presented a lightweight probabilistic broadcast algorithm, and 
analysed the evolution of processes that gossip one message. The states of the associated Markov 
chain are the number of processes that propagate one gossip message. From the designed Markov chain, 
the authors computed the distribution of the gossiping nodes. Their analysis has shown that the 
expected number of rounds to propagate the message to the entire system does not depend on the out-degree 
of nodes. These results are based on the analysis assumption that the individual out-degrees are 
uniform. However, this simplification has shown to be valid only for small systems (cf. \cite{steen2007.15}).

Bonnet \cite{FB06} studied the evolution of the in-degree distribution 
of nodes executing the Cyclon protocol \cite{VoulgarisGS05}. The states of the associated 
Markov chain are the fraction of nodes with a specific in-degree 
distribution. From the designed Markov chain the author determined the 
distribution to which the protocol converges.

There are a number of theoretical results on gossip protocols, targeted to a distributed 
aggregation.

Boyd et al. \cite{BGPS05} studied the averaging problem and analysed a gossip protocol 
in which nodes compute the average of their local measurements. The Markov chain is defined by 
a weighted random walk on the graph. Every time step, a pair of nodes (connected by an edge) 
communicates with a transition probability, and sets their values equal to the average of 
their current values. A state of the associated Markov chain is a vector of values at the end of 
the time step. The authors considered the optimization of the neighbour selection probabilities 
for each node, to find the fastest-mixing Markov chain (for fast convergence of the algorithm) 
on the graph. 

Jelasity et al. \cite{1082470} proposed a solution for aggregation in large dynamic networks, supported by
a performance analysis of the protocol. A state of the system is represented by 
a vector, the elements of which correspond to the values at the nodes, a target value of the protocol
calculated from the vector elements, and a measure of homogeneity characterizing the quality of 
local approximations. The vector evolves at every step of the system according to some 
distribution. In the analysis, the authors considered different strategies (e.g., neighbour selection) 
to optimize the protocol implementation, and calculated the expected values for the abovementioned 
protocol parameters.

Deb et al. \cite{1148678} studied the adaptation of random network coding to gossip protocols. 
The authors analysed the expected time and message complexity of two gossip protocols for message transmission 
with pure push and pure pull communication models.

%% file: protocol.tex
\section{A Gossip-based Protocol for Wireless Networks}
\label{sec:protocol}

This section describes the shuffle protocol introduced in~\cite{GVS06}. It is a gossip protocol to disseminate small data items of general interest to a collection of wireless devices. The protocol relies on replication to ensure the availability of data items in the face of dynamic behaviour, which is characteristic of wireless environments. 

The system consists of a collection of wireless nodes, each of which contributes a limited amount of storage space (which we will refer to as the node's cache) to store data items. The nodes periodically swap (shuffle) data items from their cache with a randomly chosen neighbour. In this way, nodes update their caches on a regular basis, allowing nodes to gradually discover new items as they are disseminated through the network. 

Items can be published by any user of the system, and are propagated through the network. While an item is a piece of information, a copy is the representation of the item in the network, and for each item several copies may exist. As items are gossiped between neighbouring nodes, replication may occur when a node has available storage space to keep a copy of an item it just gossiped to a neighbour.

\subsection{Protocol assumptions}
\label{sec:assum}
All nodes have a common agreement on the frequency of gossiping. However, there is no agreement on when to gossip.

In terms of storage space, we assume that all nodes dedicate the same amount of storage space to keep items locally, and that all items are of the same size. Therefore, we say that each node has a cache size of $c$. When shuffling, each node sends a fixed number $s$ of the $c$ items in the cache.

The gossip exchange is performed as an atomic procedure, meaning that once a node initiates an exchange with another node, these pair of nodes cannot become involved in another exchange until the current exchange is finished.

\subsection{Description}
\label{sec:alg}
Nodes executing the shuffle protocol initiate a shuffle periodically. In order to execute the protocol, the initiating node needs to contact a gossiping partner. We describe the protocol from the point of view of each participating node. We refer to \cite{GVS06} for a more detailed description.

Node $A$ initiates the shuffle by executing the following steps:
\vspace{-2mm}
\begin{enumerate}
	\item picks a neighbouring node $B$ at random; 
	\item selects randomly $s$ items from the local cache, and sends a copy of these items to $B$;
	\item receives $s$ items from the local cache of $B$;
 	\item checks whether any of the received items are already in its cache; if so, these received items are eliminated;
	\item adds the rest of the received items to the local cache; if the total number of items exceeds cache size $c$, removes items among the ones that were sent by $A$ to $B$, but not those that were also received by $A$ from $B$, until the cache contains $c$ items.
\end{enumerate}

In response to being contacted by $A$, node $B$ executes the following steps:
\vspace{-2mm}
\begin{enumerate}
 \item receives $s$ items from the local cache of $A$;
 \item selects randomly $s$ items from its local cache, and sends a copy of these items to $A$;
 \item checks whether any of the received items are already in its cache; if so, these received items are eliminated;
 \item adds the rest of the received items to the local cache; if the total number of items exceeds cache size $c$, removes items among the ones that were sent by $B$ to $A$, but not those that were also received by $B$ from $A$, until the cache contains $c$ items.
\end{enumerate}

According to the protocol, each node agrees to keep the items received from a neighbour. Given the limited 
storage space available in each node, keeping the items received during an exchange implies discarding some items that 
the node has in its cache. By picking the items to be discarded from the ones that have been sent to the neighbour, the 
conservation of data in the network is ensured.

\subsection{Properties}\label{sec:properties}

We are interested in characteristics of the dissemination of data items when the protocol is executed at a large scale, i.e. with a large set of nodes. For this reason, we focus on two properties that can be observed in large deployments: i) the number of replicas of an item in the network, and ii) the coverage achieved by an item over time.
\vspace{-4mm}
\subsubsection{Replication}
This property is defined as the fraction of nodes that hold a copy of a generic item $d$ in their cache, at a given moment. After an item is introduced into the network, with every shuffle involving a node that has the item in its cache, there is a chance that a new copy of the item will be created, or that the item will be discarded. As a result, with every passing round the number of copies in the network for a particular item fluctuates. Given that the storage space at the nodes is limited, items are in constant competition to place copies in the network. Since competition is fair (all items have the same chance of being replicated or discarded), eventually the storage capacity is evenly divided between the existing items. To be more precise, consider a network of $N$ nodes, in which $n$ different items have been published in total. Since there are $N\cdot c$ cache entries in the network in total, the average number of copies that an individual item has in the network will converge to $\frac{N \cdot c}{n}$. So replication will converge to $\frac{c}{n}$.
\vspace{-4mm}
\subsubsection{Coverage}
This property is defined as the fraction of nodes in the network that have seen a generic item $d$ since it was introduced into the network. As explained earlier, several copies of an item are generated after the item is first published. Due to the periodic nature of the protocol, these copies continually move through the network. This results in nodes discovering item $d$ over several rounds. With each passing round, more nodes will have seen $d$. Eventually, $d$ will have been seen by all nodes (i.e., the coverage is equal to 1). The speed at which the coverage grows is influenced by several factors (as will be explained later on) including the number of different items in the network (i.e. competition), cache size, and the size of the exchange buffer.

%% file: analysis.tex
\section{An Analytical Model of Information Dissemination}
\label{sec:analysis}

We analyse dissemination of a generic item $d$ in a network in which the nodes execute the shuffling protocol.

\subsection{Probabilities of state transitions}

\begin{wrapfigure}[17]{r}{55mm}
 \begin{center}
  \vspace{-30pt}
\includegraphics{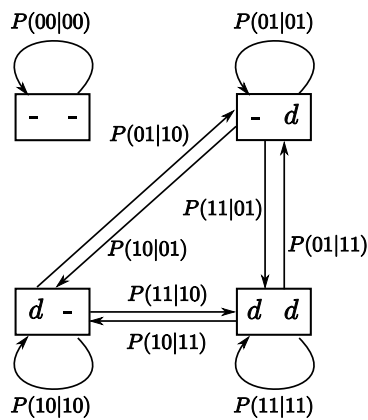} 
 \end{center}
  \vspace{-20pt}
 \caption{Symbolic representation for caches of gossiping nodes.}
 \label{fig:allsc}
\end{wrapfigure}
We present a model of the shuffle protocol that captures the presence or absence of a generic item $d$ after shuffling of two nodes $A$ and $B$. There are four possible states of the caches of $A$ and $B$ before the shuffle: both hold $d$, either $A$'s or $B$'s cache holds $d$, or neither cache holds $d$.

We use the notation $P(a_2 b_2|a_1 b_1)$ for the probability that from state $a_1 b_1$ after a shuffle we get to state $a_2 b_2$, with $a_i,b_i\in\{0,1\}$. The indices $a_1$, $a_2$ and $b_1$, $b_2$ indicate the presence (if equal to $1$) or the absence (if equal to $0$) of a generic item $d$ in the cache of an initiator $A$ and the contacted node $B$, respectively. For example, $P(01|10)$ means that node $A$ had $d$ before the shuffle, which then moved to the cache of $B$, afterwards. Due to the symmetry of information exchange between nodes $A$ and $B$ in the shuffle protocol, $P(a_2 b_2|a_1 b_1)=P(b_2 a_2|b_1 a_1)$.

Fig.~\ref{fig:allsc} depicts all possible outcomes for the caches of gossiping nodes as a state transition diagram. If before the exchange $A$ and $B$ do not have $d$ ($a_1 b_1 = 00$), then clearly after the exchange $A$ and $B$ still do not have $d$ ($a_2 b_2 = 00$). Otherwise, if $A$ or $B$ has $d$ ($a_1 = 1 \vee b_1 = 1$), the shuffle protocol guarantees that after the exchange $A$ or $B$ still has $d$ ($a_2 = 1 \vee b_2 = 1$). Therefore, the state $(-,-)$ has a self-transition, and no other outgoing or incoming transitions.

We determine values for all probabilities $P(a_2 b_2|a_1 b_1)$. They are expressed in terms of probabilities $\probsn$ and $\probdn$. The probability $\probsn$ expresses the chance of an item to be selected by a node from its local cache when engaged in an exchange. The probability $\probdn$ represents a probability that an item which can be overwritten (meaning it is in the exchange buffer of its node, but not of the other node in the shuffle) is indeed overwritten by an item received by its node in the shuffle. Due to the symmetry of the protocol, these probabilities are the same for both initiating and contacted nodes. In Sec.~\ref{sec:expl}, we will calculate $\probsn$ and $\probdn$. We write $\probnotsn$ for $1-\probsn$ and $\probkn$ for $1-\probdn$.

\begin{scenario}[$a_1 b_1 =00$] Before shuffling, neither node $A$ nor node $B$ have $d$ in their cache.
\vspace{-2mm}
\begin{description}
 \item[$a_2 b_2=00$:] neither node $A$ nor node $B$ have item $d$ after a shuffle because neither of them had it in the caches before the shuffle: $P(00|00)=1$
\item[$a_2 b_2 \in \{01,10,11\}$:] cannot occur, because none of the nodes have item $d$.
 \end{description} 
\end{scenario}

\begin{scenario}[$a_1 b_1 =01$] Before shuffling, a copy of $d$ is only in the cache of node $B$.
\vspace{-2mm}
\begin{description}
 \item[$a_2 b_2=01$:] node $A$ does not have $d$ because node $B$ had $d$ but did not select it (to send) and, thus, $B$ did not overwrite $d$, i.e. the probability is $P(01|01)= \probnotsn$
 \item[$a_2 b_2=10$:] only node $A$ has $d$ because node $B$ selected $d$ and dropped it; that is, the probability is $P(10|01)=\probsn \cdot \probdn$
 \item[$a_2 b_2=11$:] both nodes $A$ and $B$ have a copy of $d$ because node $B$ selected $d$ and kept it; that is, \(P(11|01)=\probsn \cdot \probkn\)
 \item[$a_2 b_2=00$:] cannot occur as completely discarding $d$ is not possible in the protocol; that is, if either nodes send an item, its partner keeps this copy as well, and if an item is not among the selected for a shuffle, the item is not replaced by another one (see Sec.~\ref{sec:alg}).
\end{description}
\end{scenario}

\begin{scenario}[$a_1 b_1=10$] Before shuffling, $d$ is only in the cache of node $A$.
Due to the symmetry of nodes $A$ and $B$, this scenario is symmetric to the previous one with $P(a_2 b_2|10)=P(b_2 a_2|01)$.
\end{scenario}

\begin{scenario}[$a_1 b_1=11$] Before shuffling, $d$ is in the cache of node $A$ as well as in the cache of node $B$.
\vspace{-1.5mm}
\begin{description}
 \item[$a_2 b_2=01$:] only node $B$ has $d$ because node $A$ selected $d$ and dropped it and node $B$ did not select $d$; that is, \( P(01|11)= \probsn \cdot \probdn \cdot \probnotsn \)
 \item[$a_2 b_2=10$:] this outcome is symmetric to the previous one:
\( P(10|11) = \probnotsn \cdot \probsn \cdot \probdn\)
 \item[$a_2 b_2=11$:] after the shuffle both nodes $A$ and $B$ have $d$, because:
 \subitem nodes $A$ and $B$ had $d$ but both did not select it, i.e. $\probnotsn \cdot \probnotsn$;
 \subitem both nodes $A$ and $B$ selected $d$ (thus, both kept it), i.e. $\probsn \cdot \probsn$;
 \subitem node $A$ selected $d$ and kept it and node $B$ did not select $d$: $\probsn \cdot \probkn \cdot \probnotsn$;
 \subitem symmetric case with the previous one: $\probnotsn \cdot \probsn \cdot \probkn$.

\noindent Thus, \(P(11|11)=\probnotsn \cdot \probnotsn + \probsn \cdot \probsn + 2 \cdot \probsn \cdot \probnotsn \cdot \probkn \)
\item[$a_2 b_2=00$:] cannot occur, discarding of an item is not permitted by the protocol (see Sec.~\ref{sec:alg}).
\end{description}
\end{scenario}

\subsection{Probabilities of selecting and dropping an item}
\label{sec:expl}

The following analysis assumes that all node caches are full (that is, the network is already running for a while). Moreover, we assume a uniform distribution of items over the network; this assumption is supported by experiments in \cite{GVS06,steen2007.15}.

\begin{figure}[!hptb]
\begin{minipage}[b]{0.5\linewidth}
\centering
\includegraphics[]{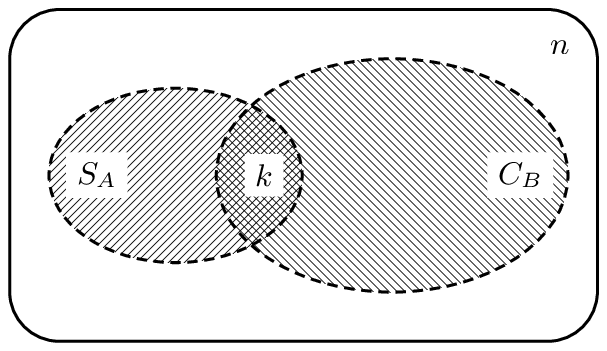}
\caption{$k$ items in $S_A \cap C_B$}
\label{fig:schemenc}
\end{minipage}%
\begin{minipage}[b]{0.5\linewidth}
\centering
\includegraphics[]{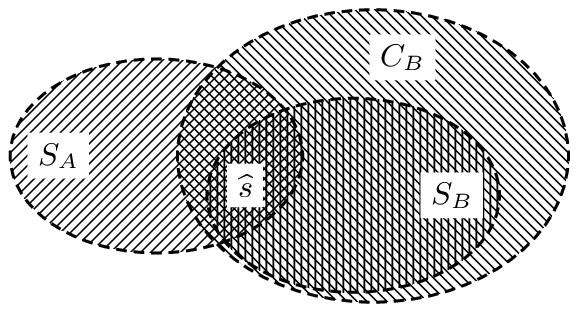}
\caption{$\sasb$ items in $S_A \cap S_B$}
\label{fig:sasbschema}
\end{minipage}
\end{figure}

Consider nodes $A$ and $B$ engaged in a shuffle, and let $B$ receive the exchange buffer $S_A$ from $A$. Let $k$ be the number of duplicates (see Fig.~\ref{fig:schemenc}), i.e. the items of an intersection of the node cache $C_B$ and the exchange buffer of its gossiping partner $S_A$ (i.e. $S_A \cap C_B$). Recall from Sec.~\ref{sec:assum} that $C_A$ and $C_B$ contain the same number of items for all $A$ and $B$, and likewise for $S_A$ and $S_B$; we use $c$ and $s$ for these values. The total number of different items in the network is denoted as $n$.

The probability of selecting an item $d$ in the cache is the probability of a single selection trial (i.e. $\frac{1}{c}$) times the number of selections (i.e. $s$): \( \probsn = \probs \). 
Thus, the probability that an item $d$ in the cache is not selected is: \( \probnotsn = 1 - \probsn = \frac{c-s}{c} \). 

Consider Figs.~\ref{fig:schemenc} and~\ref{fig:sasbschema}. The shuffle protocol demands that all items in $S_A$ are kept in $C_B$ after the shuffle. This implies that: a) all items in $S_A \sminus C_B$ will overwrite items in $S_B \subseteq C_B$, and b) all items in $S_A \cap C_B$ are kept in $C_B$. Thus, the probability that an item from $S_B$ will be overwritten is determined by the probability that an item from $S_A$ is in $C_B$, but not in $S_B$. Namely, the items in $S_B \sminus S_A$ provide a space in the cache for items from $S_A \sminus C_B$. We would like to express the probability $\probdn$ of a selected item $d$ in $S_B \sminus S_A$ (or $S_A \sminus S_B$) to be overwritten by another item in $C_B$ (or $C_A$). Due to symmetry, this probability is the same for $A$ and $B$; therefore, we only calculate the probability that an item in $S_B \sminus S_A$ is dropped from $C_B$. The expected value of this probability depends on how many duplicates a node receives from its gossiping partner:
\[E[\probdn] =  \begin{cases}
\bigsum{k=0}{s}(\probdwk \cdot \probok) &\text{if } s + c \loeq n\\
\bigsum{k=(s+c)-n}{s}(\probdwk \cdot \probok) &\text{otherwise}
\end{cases}
\]
where $\probok$ is the probability of having exactly $k$ items in $S_A \cap C_B$, and $\probdwk$ is the probability that an item in $S_B \sminus S_A$ is dropped from $C_B$ given $k$ duplicates in $S_A\cap C_B$. The case distinction is because if $s + c > n$, then clearly there are at least $(s+c) -n$ items in $S_A \cap C_B$.

From the $\binom{n}{s}$ possible sets $S_A$, we compute how many have $k$ items in common with $C_B$. Firstly, there are $\binom{c}{k}$ ways to choose $k$ such items in $C_B$. Secondly, there are $\binom{n-c}{s-k}$ ways to choose the remaining $s-k$ items outside $C_B$. So in total, $\binom{c}{k} \cdot \binom{n-c}{s-k}$ possible sets $S_A$ have $k$ items in common with $C_B$. Hence, under the assumption of a uniform distribution of the data items over the caches of the nodes,\footnote{Here we use a generalization of the usual definition of binomial coefficients to negative integers. That is, for all $m$ and $l \geq 0$, $\binom{m}{l} = (-1)^l \binom{-m+l-1}{l}$ (cf.~\cite{HHP97})}
\(\probok = \binom{c}{k} \frac{\binom{n-c}{s-k}}{\binom{n}{s}}\).
The expected value of $\probdwk$ is:
\[E[\probdwk] =  \begin{cases}
\bigsum{\sasb=0}{k}\probdwsh \cdot \probosh &\text{ if } s + k \loeq c\\
\bigsum{\sasb=(s+k)-c}{k}\probdwsh \cdot \probosh &\text{otherwise}
\end{cases}
 \]
where $\sasb$ is the number of items in $S_A \cap S_B$ (see Fig.~\ref{fig:sasbschema}). The case distinction is because if $s + k > c$ (with $k$ the number of items in $S_A \cap C_B$), then clearly there are at least $(s+k) - c$ items in $S_A \cap S_B$.

Among the $s$ items in $S_B$, there are $\sasb$ items also in $S_A$, and thus only the $s - \sasb$ items in $S_B \sminus S_A$ can be dropped from $C_B$. $\probdwsh$ is the probability that an item in $S_B \sminus S_A$ is dropped from $C_B$, given $\sasb$ items in $S_A \cap S_B$:
\[\probdwsh =  \begin{cases}
0 &\text{if } s = \sasb \\
\frac{s-k}{s-\sasb} &\text{otherwise}
\end{cases}
\]
$\probosh$ is the probability of having exactly $\sasb$ items in $S_A\cap S_B$: 
\(E[\probosh] = \binom{s}{\sasb} \frac{\binom{c-s}{k-\sasb}}{\binom{c}{k}}\).
The intuition behind this expected value of $\probosh$ is similar to the one of $\probok$. From the $\binom{c}{k}$ possible sets $S_A$, we compute how many have $\sasb$ items in common with $S_B$. That is, there are $\binom{s}{\sasb}$ ways to choose $\sasb$ items in $S_B$, and $\binom{c-s}{k-\sasb}$ ways to choose the remaining $k-\sasb$ items outside $S_B$.

Let's assume $2s \leq c \leq n-s$ (because then $s+c \leq n$ and $s+k \leq 2s \leq c$). Then, substituting in the expression for $E [\probdn]$ in case  $s+c \leq n$, and noting that in the summand $k=s$ the factor $\probdwshp{s}$ is equal to zero, we get:
\begin{eqnarray}
E [\probdn] &~=~& \bigsum{k=0}{s-1} \binom{c}{k} \frac{\binom{n-c}{s-k}}{\binom{n}{s}}  
    \bigsum{\sasb=0}{k} \frac{s-k}{s-\sasb} 
\binom{s}{\sasb} \frac{\binom{c-s}{k-\sasb}}{\binom{c}{k}}  \nonumber \\
&~=~& \frac{n-c}{\binom{n}{s}}\bigsum{k=0}{s-1} \binom{(n-c)-1}{(s-k)-1} \bigsum{\sasb=0}{k} \frac{\binom{c-s}{k-\sasb} \binom{s}{\sasb}}{s-\sasb}
\label{eq:exact}
\end{eqnarray}
The probability of keeping an item $d$ in $S_B \sminus S_A\! \subseteq\! C_B$ can be expressed as $\probkn = 1 - \probdn$.

\subsection{Simplification of $\probdn$}
\label{subsec:simplifiedPdrop}
In order to gain a clearer insight into the emergent behaviour of the gossiping protocol we make an effort to simplify the formula for the probability $\probdn$ of an item in $S_B \sminus S_A$ to be dropped from $C_B$ after a shuffle. Therefore, we re-examine the relationships between the $k$ duplicates received from a neighbour, the $\sasb$ items of the overlap $S_A \cap S_B$, and $\probdn$. Let's estimate $\probdwk$ by considering each item from $S_A$ separately, and calculating the probability that the item is a duplicate (i.e., is also in $C_B$). The probability of an item from $S_A$ to be a duplicate (also present in $C_B$) is $\frac{c}{n}$. In view of the uniform distribution of items over the network, the items in a node's cache are a random sample from the universe of $n$ data items; so all items in $S_A$ have the same chance to be a duplicate. Thus, the expected number of items in $S_A \cap C_B$ can be estimated by \( E[k] = s \cdot \frac{c}{n} \). And the expected number of items in $S_A \cap S_B$ can be estimated by $E[\sasb] = k \cdot \frac{s}{c}$, because only the $k$ items in $S_A\cap C_B$ may end up in $S_A\cap C_B$; $\frac{s}{c}$ captures the probability that an item from $C_B$ is also selected to be in $S_B$.
It follows that the probability of an item in $S_B \sminus S_A$ to be dropped from $C_B$ after the shuffle is
\(E[\probdn] = \frac{s-k}{s-\sasb} = \frac{s - s \cdot \frac{c}{n}}{s - s \cdot \frac{c}{n} \cdot \frac{s}{c}} = \frac{n-c}{n-s}.
\)
The complementary probability of keeping an item is 
\( E[\probkn] = 1 - \frac{n-c}{n-s} = \frac{c-s}{n-s} \). These estimates are valid for general $s \leq c \leq n$.

Substituting the expressions for $\probsn$ and the simplified $\probdn$ into the formulas for the transition probabilities in Fig.~\ref{fig:allsc}, we obtain:
\[
\begin{array}{rclrcl}
P(01|01)=P(10|10) &=& \frac{c-s}{c} & P(01|11)=P(10|11) &=& \frac{s}{c} \frac{c-s}{c} \frac{n-c}{n-s} \vspace{2mm}\\
P(10|01)=P(01|10) &=& \frac{s}{c} \frac{n-c}{n-s} \hspace*{1.5cm} & P(11|11) &=& 1 - 2 \frac{s}{c} \frac{c-s}{c} \frac{n-c}{n-s} \vspace{2mm}\\
P(11|01)=P(11|10) &=& \frac{s}{c} \frac{c-s}{n-s}
\end{array}
\]

\begin{figure}[!hptb]
\vspace{-4mm}
\begin{minipage}[b]{0.5\linewidth}
\centering
\includegraphics[width=1.0\textwidth]{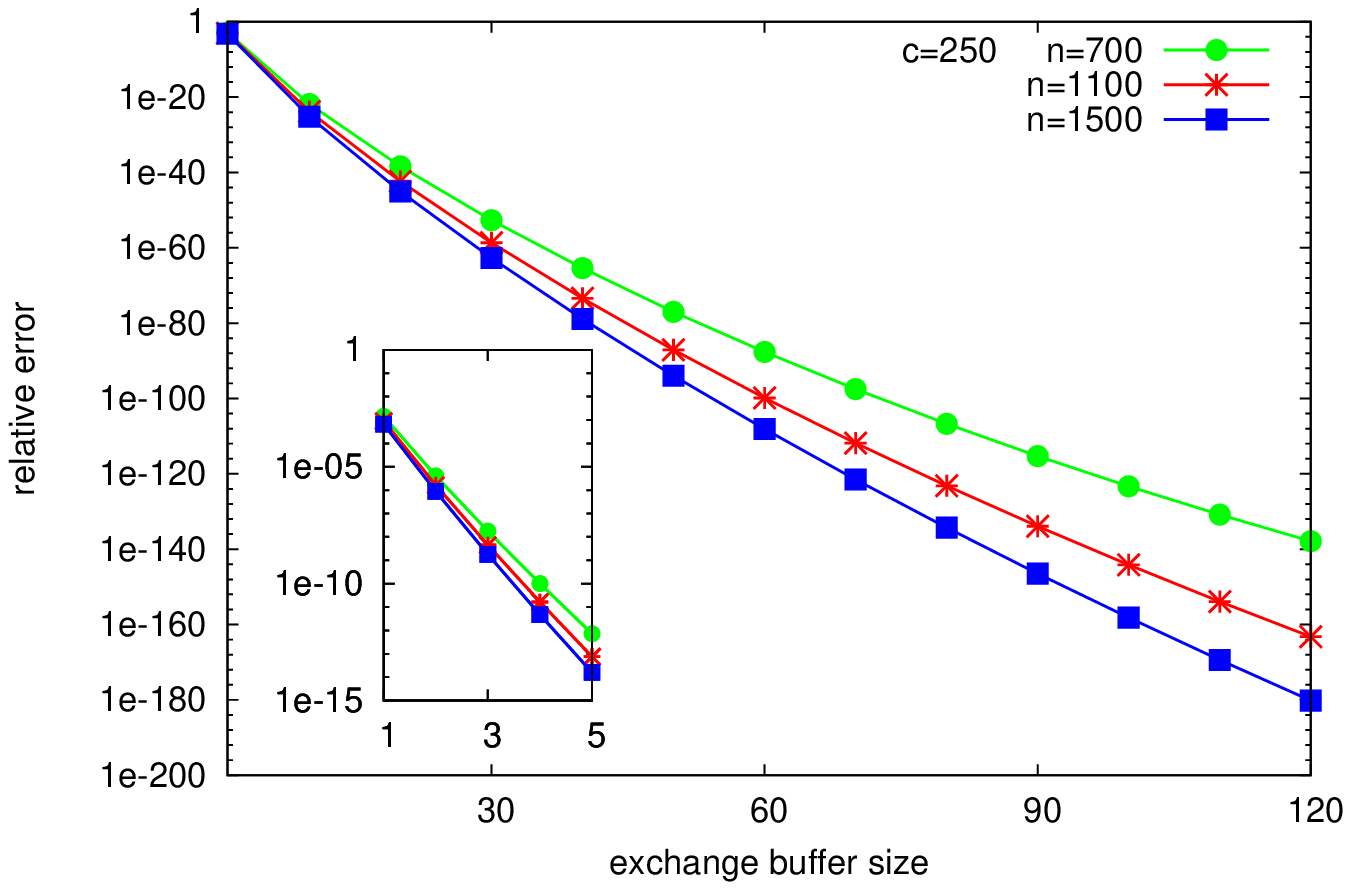}
\end{minipage}%
\begin{minipage}[b]{0.5\linewidth}
\centering
\includegraphics[width=1.0\textwidth]{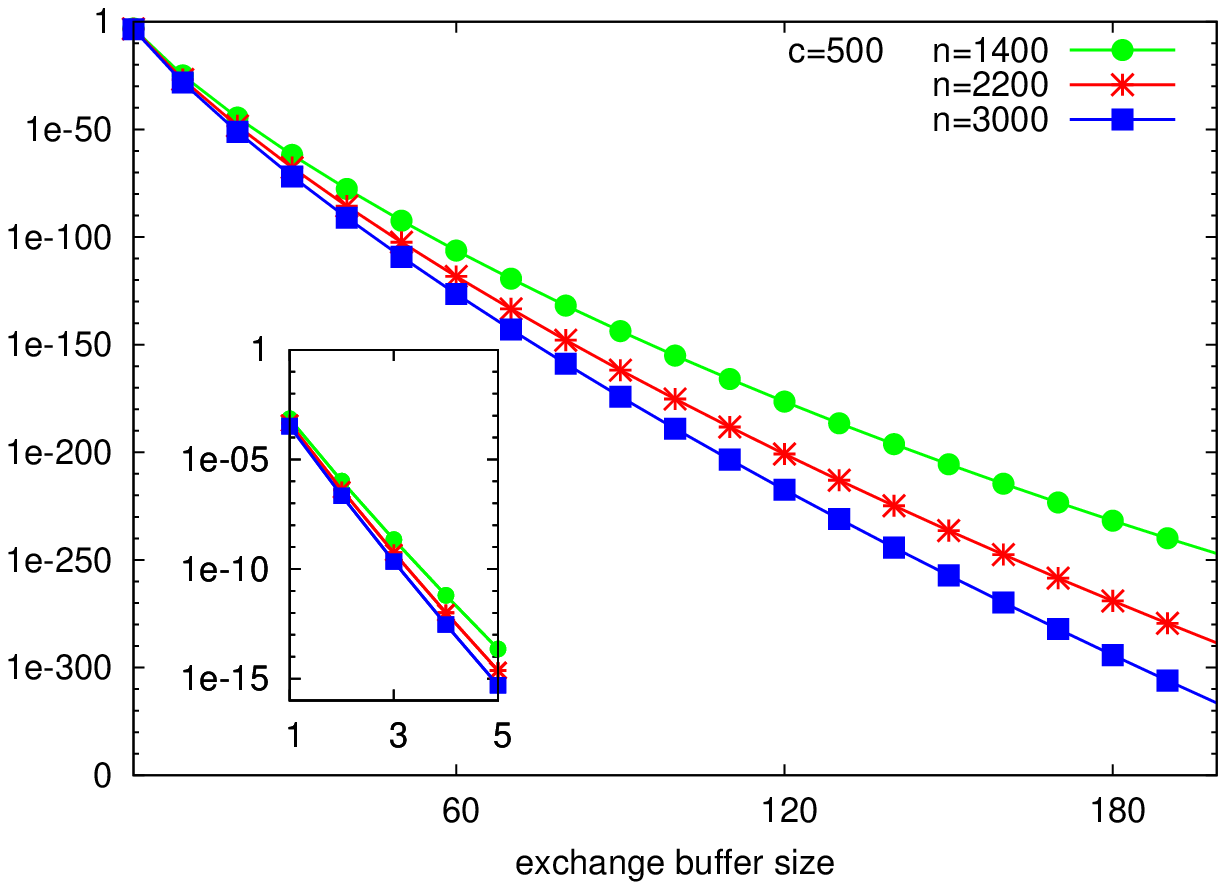}
\end{minipage}
\caption{The difference of the accurate $\probdn$ and its approximation, for different values of $n$ and $c$.}
\vspace{-3mm}
\label{fig:pdrop_comparison1}
\end{figure}

In order to verify the accuracy of the proposed simplification for $E[\probdn]$, we compare the simplification and the accurate formula \eqref{eq:exact} for different values of $n$. We plot the difference of the accurate $\probdn$ and the simplification, for cache sizes $c=250$ and $c=500$ (Fig.~\ref{fig:pdrop_comparison1}).

\subsection{Correction factor}
We now examine how closely the simplified formula $E[\probdn] = \frac{n-c}{n-s}$ (here referred as $S(n,c,s)$) approximates formula \eqref{eq:exact} (here referred as $E(n,c,s)$). We compared the difference between these two formulas using an implementation on the basis of common fractions, which provides loss-less calculation \cite{bigj}. We observed that the inverse of the difference of the inverse values of both formulas, i.e. $e_{c,s}(n) = \left( E(n,c,s)^{-1} - S(n,c,s)^{-1} \right)^{-1}$, exhibits a certain pattern for different values of $n$, $c$ and $s$. For $s=1$, $E(n,c,1) = \frac{n-c}{n}$, whereas $S(n,c,1) = \frac{n-c}{n-1}$. We then   investigate the correction factor $\theta$ in $E(n,c,s) = \frac{n-c}{(n-s) + \theta }$. Thus, for $s=1$ we have $\theta=1$. Yet, for $s > 1$ the situation turned out to be more complicated. For $s=2$, we got $e_{4,2}(7) - e_{4,2}(6) = 3.5$, $e_{4,2}(8) - e_{4,2}(7) = 4$, $e_{4,2}(9) - e_{4,2}(8) = 4.5$, and etc. Therefore we calculated the first, the second and other (forward) differences\footnote{A forward difference of discrete function $f: \mathbb{Z}\rightarrow\mathbb{Z}$ is a function $\Delta f:\mathbb{Z}\rightarrow\mathbb{Z}$ with $\Delta f(n)=f(n+1)-f(n)$ (cf. \cite{AS72}).} over $n$. We recognized that the $s$-th difference of the function $e_{c,s}(n)$ is always $\frac{1}{s}$. Moreover, at the point $n=0$ the $1$st, \ldots, $s$-th differences of the function $e_{c,s}$ exhibit a pattern similar to the Pascal triangle \cite{GKP94}; i.e. for $d \ge 1$ the $d$-th difference is: $({\rm \Delta}^d\; e_{c,s})(0) = \frac{1}{s \cdot \binom{s-1}{d}}$ (assuming $\binom{a}{b}=0$, whenever $b > a$). Knowing the initial difference at the point $n=0$, we were able to use the Newton forward difference equation \cite{AS72} to derive the following formula for $n>0$: $E[\probdn] = \frac{n-c}{(n-s) + \frac{1}{\corr}}$, where

\begin{equation}
\corr 
~=~ \bigsum{d=0}{s-1} \frac{\binom{n}{d}}{s \cdot \binom{s-1}{d}} 
~=~ \frac{\binom{n}{s}}{(n-s)+1} \cdot \bigsum{d=0}{s-1} \frac{ 1 }{\binom{n-d}{(s-1)-d}} 
\label{eq:simpwcor}
\end{equation}
In this equation the sum is finite because due to the observation that the $s$-th difference is constant $\frac{1}{s}$, all higher differences are $0$.

Extensive experiments with Mathematica and Matlab indicate that $\frac{n-c}{(n-s) + \frac{1}{\corr}}$ and formula \eqref{eq:exact} coincide. We can also see from Fig.~\ref{fig:pdrop_comparison1} that the correction factor is small.

\subsection{Optimal size for the exchange buffer}
\label{sec:optimal-s}

\begin{wrapfigure}[14]{r}{0.52\textwidth}
 \vspace{-30pt}
 \begin{center}
\includegraphics[scale=.50]{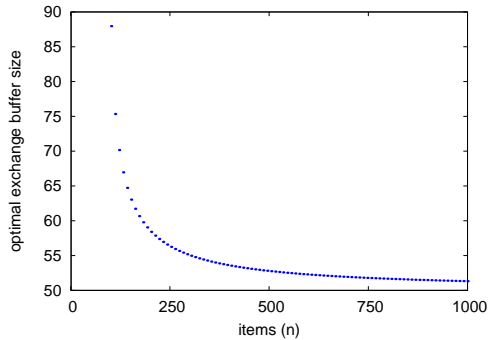}
 \end{center}
  \vspace{-20pt}
 \caption{Optimal value of exchange buffer size, depending on $n$.}
 \label{fig:optim}
\end{wrapfigure}
We study what is the optimal value for fast convergence of replication and coverage with respect to an item $d$.
Since $d$ is introduced at only one node in the network, one needs to optimize the chance that an item is duplicated.
That is, the probabilities $P(11|01)$ and $P(11|10)$ should be optimized (then $P(01|11)$ and $P(10|11)$ are optimized
as well, intuitively because for each duplicated item in a shuffle, another item must be dropped).
These probabilities both equal $\frac{s}{c}\frac{c-s}{n-s}$; we compute when the $s$-derivative of this formula is zero.
This yields the equation $s^2-2ns+nc=0$; taking into the account that $s\leq n$, the only solution of this equation is
$s = n - \sqrt{n(n-c)}$.  We conclude that this is the optimal value for $s$ to obtain fast convergence of replication
and coverage. This will also be confirmed by the experiments and analyses in the following sections.

%% file: evaluation.tex
\section{Experimental Evaluation}
\label{sec:expeval}

In order to test the validity of the analytical model of information spread under the shuffle protocol presented in the previous section, we followed an experimental approach. We compared properties observed while running the shuffle protocol in a large-scale deployment with simulations of the model under the same conditions. 
These experiments show that the analytical model indeed captures information spread of the shuffle protocol. 
We note that a simulation of the analytical model is much more efficient (in computation time and memory 
consumption) than a simulation of the implementation of the shuffle protocol.

The experiments simulate the case where a new item $d$ is introduced at one node in a network, in which all caches are full and uniformly populated by $n=500$ items. They were performed on a network of $N=2500$ nodes, arranged in a square grid topology (50$\times$50), where each node can communicate only with its four immediate neighbours (to the North, South, East and West). 

This configuration of nodes is arbitrary, we only require
a large number of nodes for the observation of emergent behaviour. Our aim
is to validate the correctness of our analytical model, not to test the endless
possibilities of network configurations. The model and the shuffle protocol
do not make any assumptions about the network. The network configuration
is provided by the simulation environment and can easily be changed into
something different, e.g. other network topology. For this reason, we have
chosen this large grid for testing, although other configurations could have 
been possible. 

Each node has a cache size of $c=100$, and sends $s$ items when gossiping. In each round, every node randomly selects one of its neighbours, and updates its state according to the transition probabilities introduced before (Fig.~\ref{fig:allsc}). This mimics (the probabilities of) an actual exchange of items between a pair of nodes according to the shuffle protocol. While in the protocol, this results in both nodes updating the contents of their caches, in a simulation using the analytical model, updating the state of a node refers to updating only one variable: whether the node is in possession of the item $d$ or not. In the experiments, after each gossiping round, we measured the total number of occurrences of $d$ in the network (replication), and how many nodes in total have seen $d$ (coverage); see Sec.~\ref{sec:properties}.

In order to fill the caches of the nodes with a random selection of items, the measurements are initiated after 1000 rounds of gossiping. In other words, 500 different items are inserted at the beginning of the simulation, and shuffled for 1000 rounds. During this time, items are replicated and the replicas fill the caches of all nodes. At round 1000, a copy of the fresh item $d$ is inserted at a random location, and its spread through the network is tracked over the next 2000 rounds.

\begin{figure}[!htpb]
\begin{minipage}[b]{0.5\linewidth}
\vspace{-3mm}
\begin{center}
\includegraphics[width=0.8\textwidth]{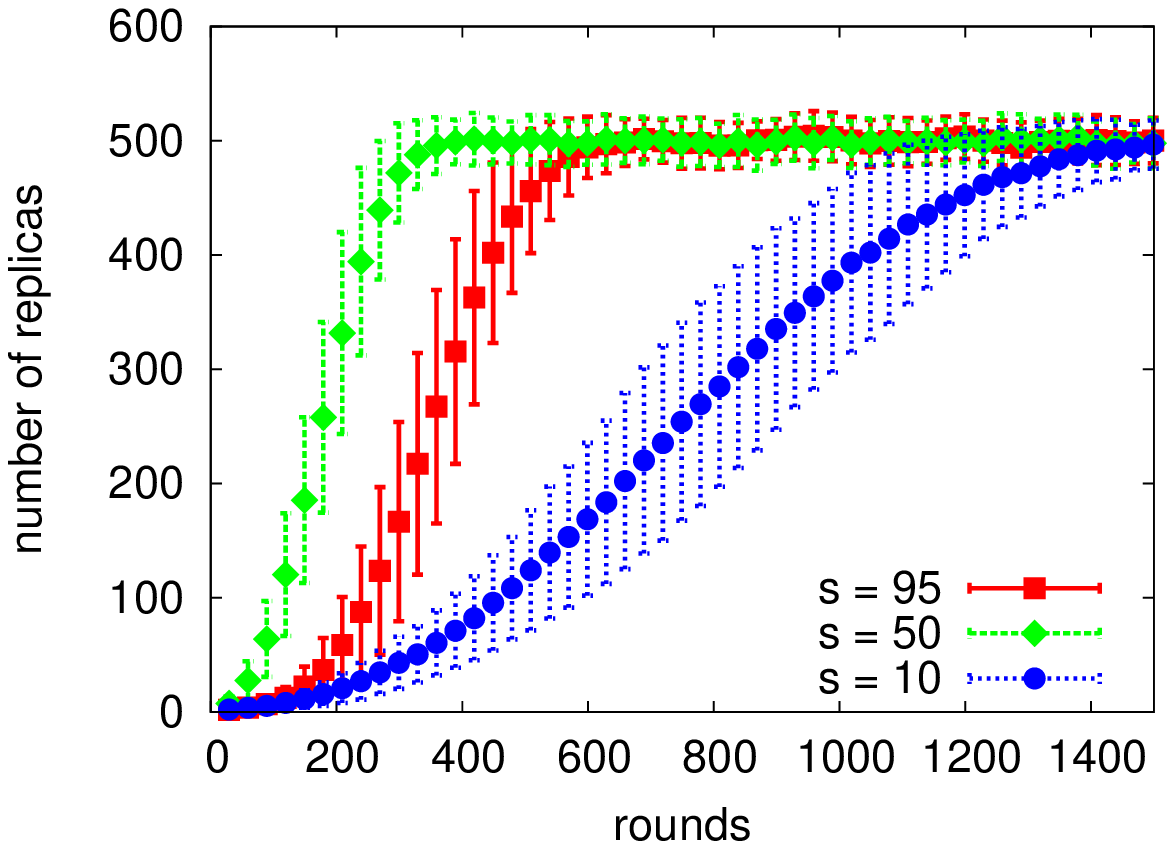}
\end{center}
\end{minipage}%
\begin{minipage}[b]{0.5\linewidth}
\vspace{-3mm}
\begin{center}
\includegraphics[width=0.8\textwidth]{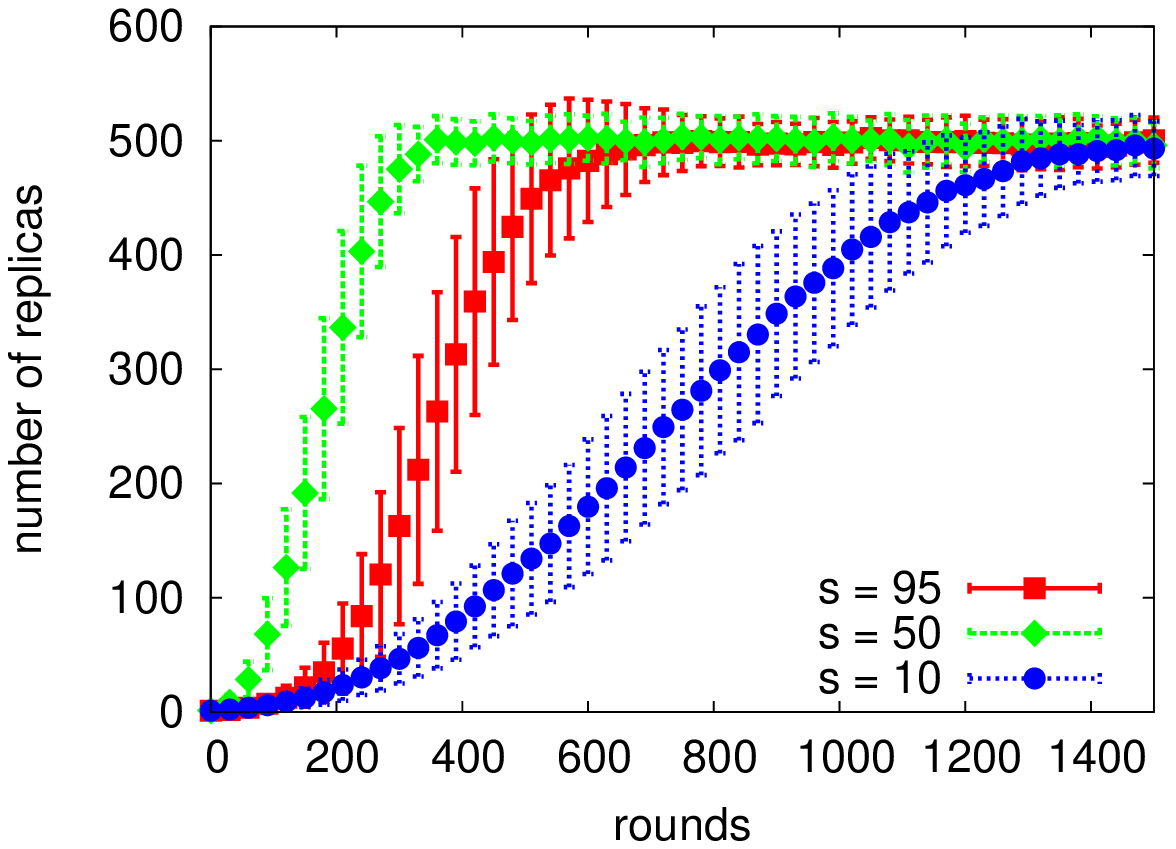}
\end{center}
\end{minipage}
\begin{minipage}[b]{0.5\linewidth}
\vspace{-1mm}
\begin{center}
\includegraphics[width=0.8\textwidth]{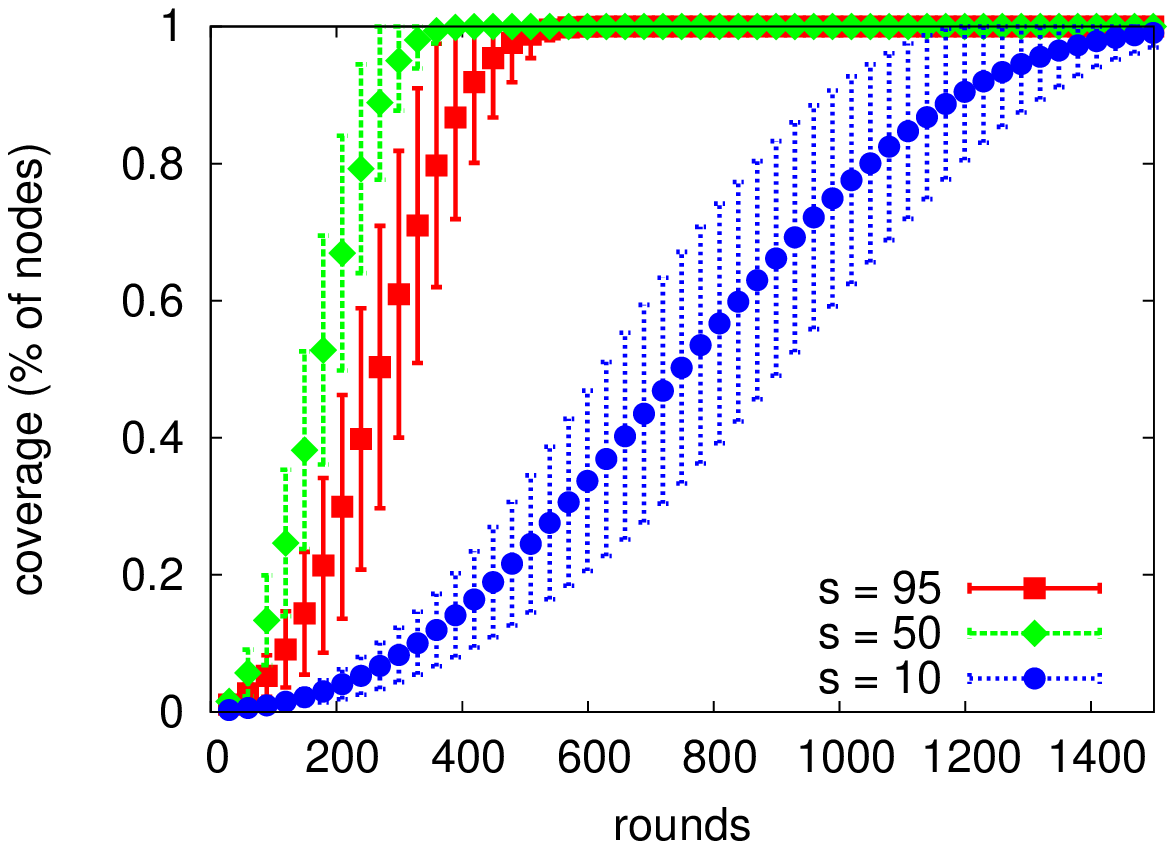}
\end{center}
\end{minipage}%
\begin{minipage}[b]{0.5\linewidth}
\vspace{-1mm}
\begin{center}
\includegraphics[width=0.8\textwidth]{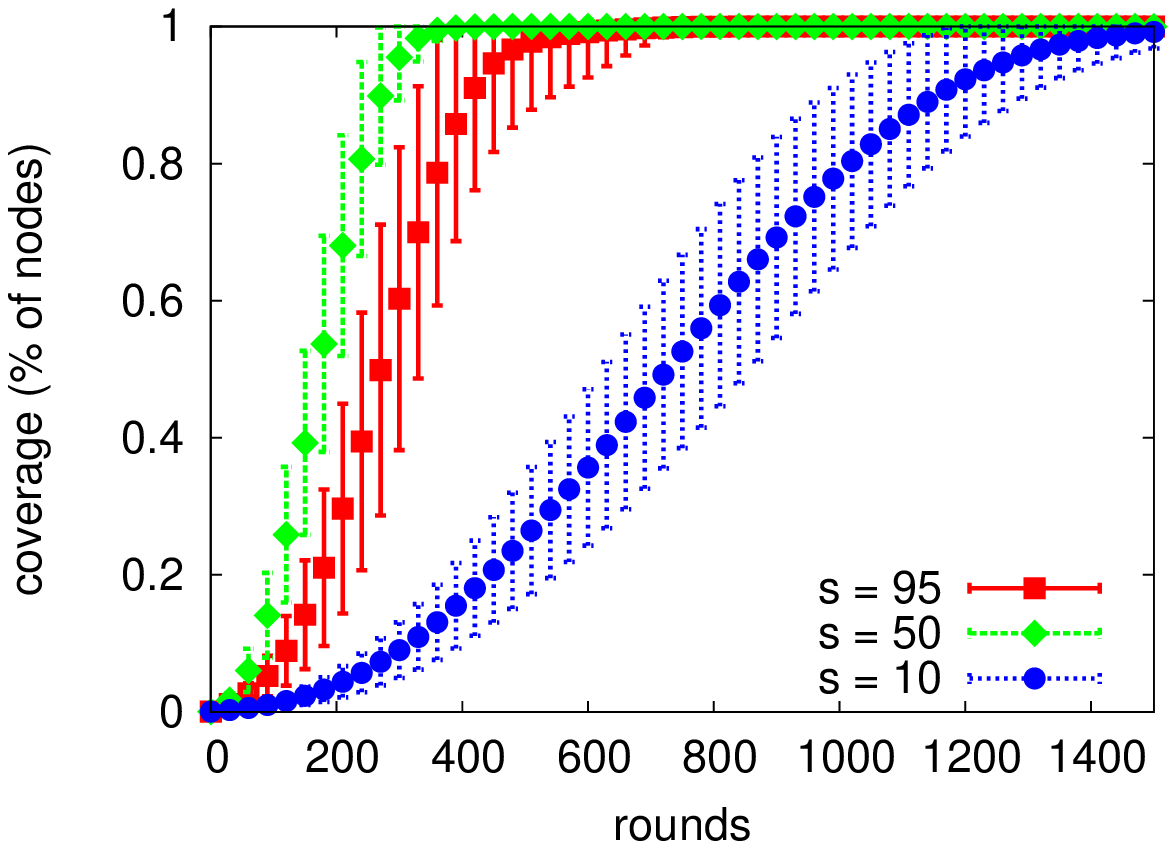}
\end{center}
\end{minipage}
\caption{The shuffle protocol (left) and the model (right), for $N=2500$,  $n=500$, $c=100$ and different values of $s$.}
\label{fig:compreplcov}
\end{figure}

Fig.~\ref{fig:compreplcov} shows the behaviour of both the shuffle protocol and the analytical model in terms of replication and coverage of $d$, for various values of $s$. Each curve in the graphs represents the average and standard deviation calculated over 100 runs. The experiments with the model calculate $\probdn$  using the simplified formula $\frac{n-c}{n-s}$ described in Sec.~\ref{subsec:simplifiedPdrop}. It can be observed very clearly that the results obtained from the model (right) resemble closely the ones from executing the protocol (left).

We note that in all cases, the network converges to a situation in which there are 500 copies of $d$, meaning that replication is $\frac{500}{2500}=0.2$; this agrees with the fact that $\frac{c}{n}=\frac{100}{500}=0.2$. Moreover, our experiments show that replication and coverage display the fastest convergence when $s=50$; this agrees with the fact that $n - \sqrt{n(n-c)}=500 - \sqrt{500\cdot 400}\approx 50$ (cf.\ Sec.~\ref{sec:optimal-s}).

%% file: modelling.tex
\section{Round-based Modelling of Protocol Properties}
\label{sec:modelling}

In this section we exploit the analytical model of information dissemination to perform
a mathematical analysis of replication and coverage with regard to the shuffle protocol.
For the particular case of a network with full connectivity, where a node can gossip with any other node 
in the network, we can find explicit expressions for the dissemination of a generic item $d$ in terms of the probabilities presented in Sec.~\ref{sec:analysis}. We construct two differential equations that capture replication and coverage of item $d$ from a round-based perspective. The advantage of this approach is that we can determine the long-term behaviour of the system as a function of the parameters.

\subsection{Replication}
\label{subsec:replication}

One node introduces a new item $d$ into the network at time $t=0$, by placing $d$ into its cache. From that moment on, $d$ is replicated as a consequence of gossiping among nodes.

Let $x(t)$ represent the percentage of nodes in the network that have $d$ in their cache at time $t$, where each gossip round takes one time unit. The variation in $x$ per time unit $\frac{d x}{dt}$ can be derived based on the probability that an item $d$ will replicate or disappear after an exchange between two nodes, where at least one of the nodes has $d$ in its cache:
\begin{eqnarray}
\frac{d x}{dt} & = & [P(11|10) + P(11|01)] \cdot (1-x) \cdot x - [P(10|11) + P(01|11)] \cdot x \cdot x \nonumber 
\end{eqnarray}

\noindent
The first term represents duplication of $d$ when a node that has $d$ in its cache initiates the shuffle, and contacts a node that does not have the item. The second term represents the opposite situation, when a node that does not have the item $d$ initiates a shuffle with a node that has $d$. The third and fourth term in the equation represent the cases where both gossiping nodes have $d$ in their cache, and after the exchange only one copy of $d$ remains. Substituting $P(11|10)=P(11|01)=\frac{s}{c}\frac{c-s}{n-s}$ and $P(10|11)=P(01|11)=\frac{s}{c}\frac{n-c}{n-s}\frac{c-s}{c}$, we obtain
\begin{eqnarray}
\frac{d x}{dt} = 2 \cdot \frac{s}{c} \cdot \frac{c-s}{n-s} \cdot x \cdot (1 - \frac{n}{c} \cdot x)
\end{eqnarray}

\noindent
The solution of this equation, taking into account that $x(0)=\frac{1}{N}$ with $N$ the number of nodes in the network, is
\begin{equation}
x(t) = \frac{e^{\alpha t}}{(N-\frac{n}{c})+\frac{n}{c}e^{\alpha t}}
\label{eq:solreplica}
\end{equation}
where $\alpha$ denotes $2\frac{s}{c}\frac{c-s}{n-s}$.
By imposing stationarity, i.e. $\frac{d x}{dt}=0$, we find the stationary solution $\frac{c}{n}$.
Hence, this calculation confirms the observation in Sec.~\ref{sec:properties} that the network converges to a situation in which replication of $d$ is $\frac{c}{n}$.

\begin{figure}[htb]
        \begin{center}                                  
        \begin{tabular}{c}
        \begin{minipage}[c]{\textwidth}
	\begin{center}
        \begin{tabular}[c]{c c}
        \includegraphics[scale=0.50]{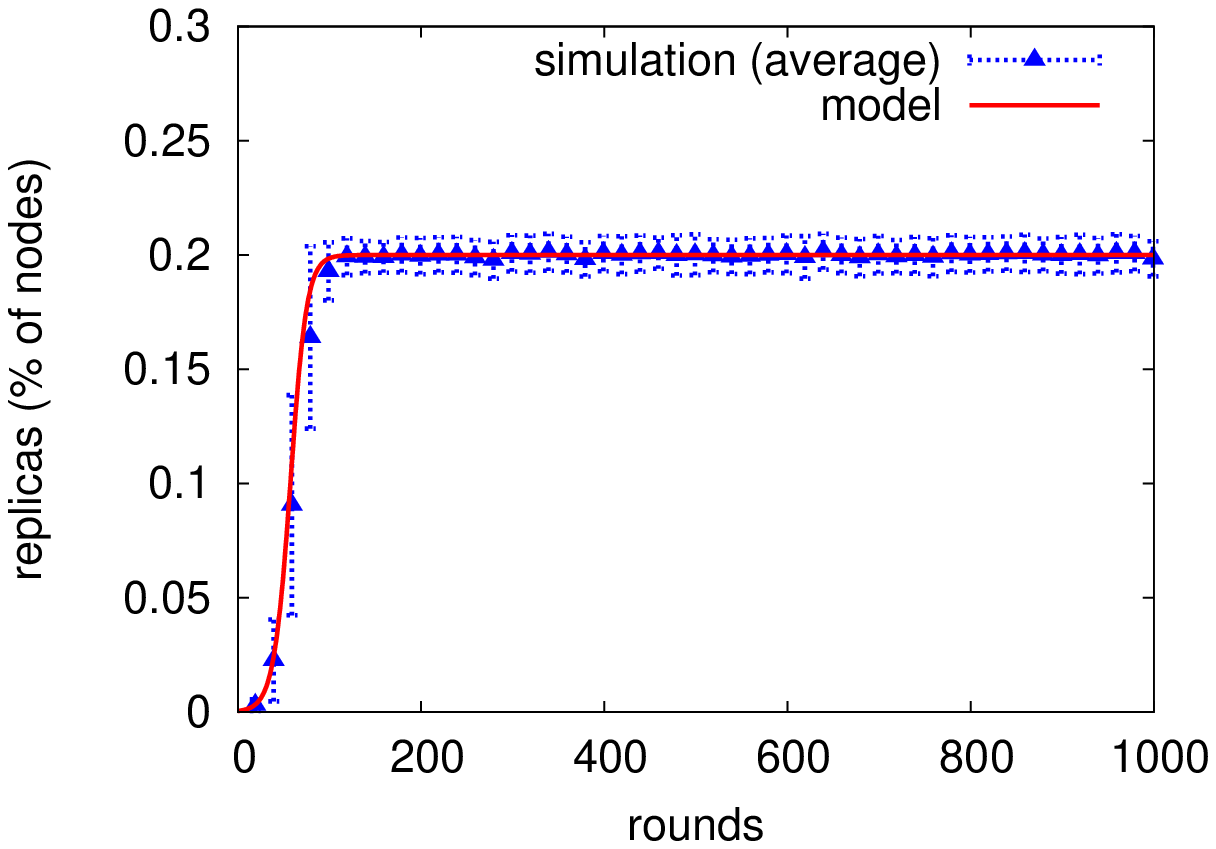} 
	& \includegraphics[scale=0.50]{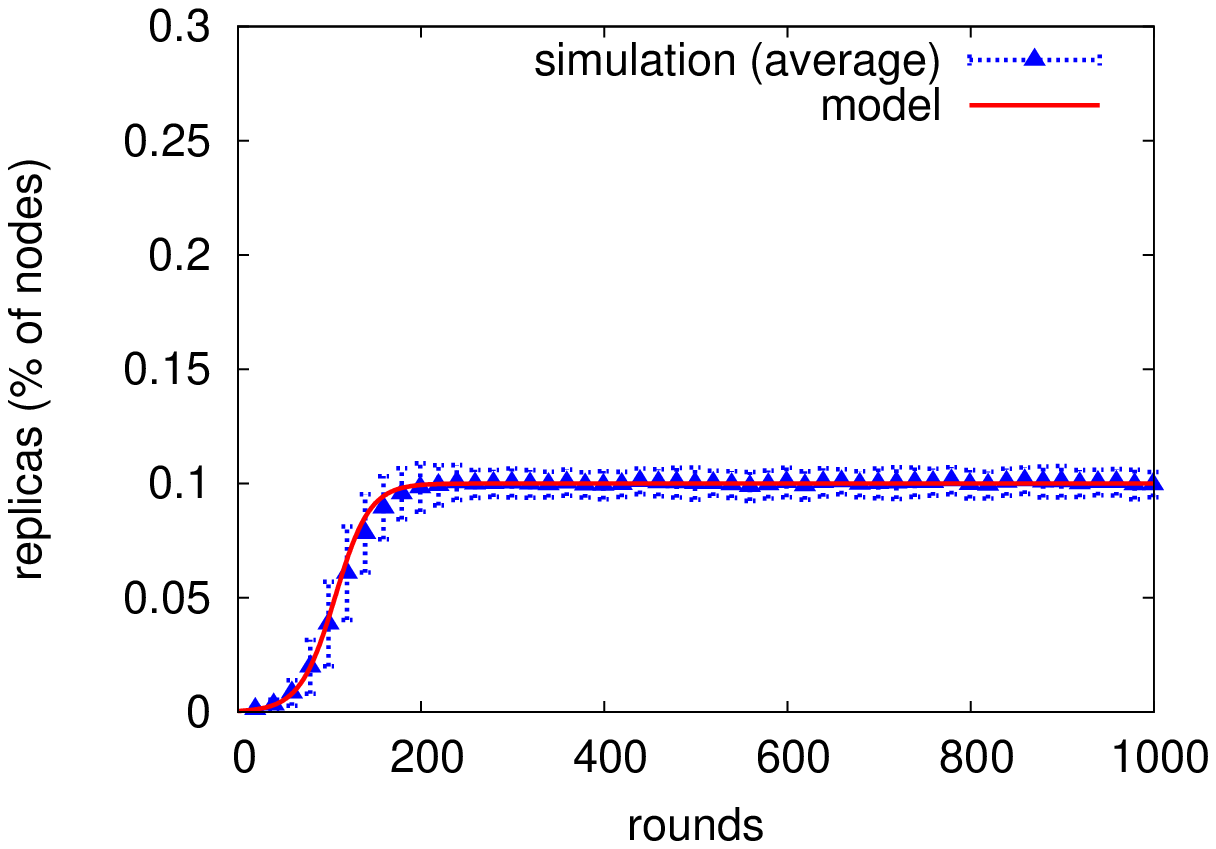} \\
	500 items & 1000 items \\
	\end{tabular} 
        \end{center}
	\end{minipage} \\
        \begin{minipage}[c]{\textwidth}
	\begin{center}
        \begin{tabular}[c]{c}
        \includegraphics[scale=0.50]{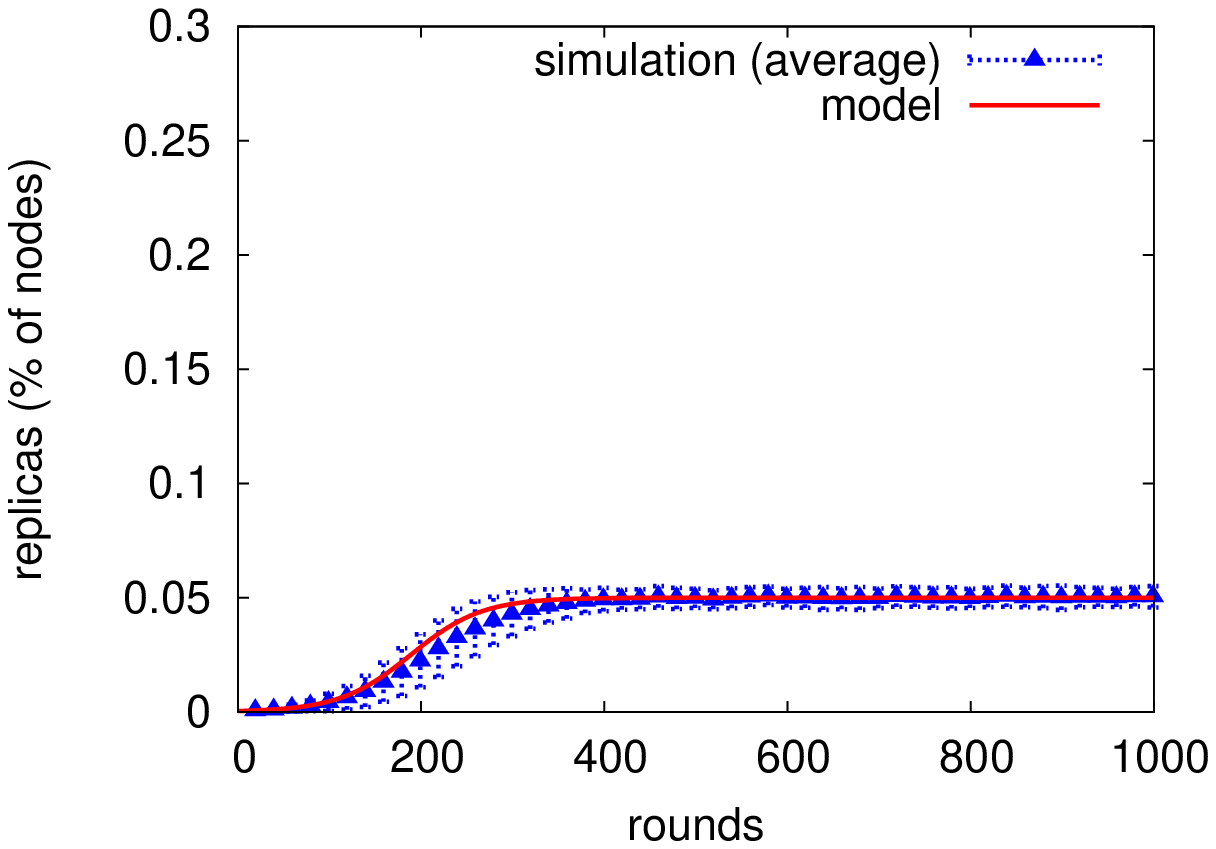} \\
	2000 items \\
	\end{tabular}
	\end{center} 
        \end{minipage} \\
        \end{tabular}
	\end{center}
        \caption{Percentage of nodes in the network with a replica of item $d$ in their cache, for $N=2500$, $c = 100$, $s = 50$, and $n=500$, $n=1000$ or $n=2000$.}
        \label{fig:replicas}
\end{figure}

We evaluate the accuracy of $x(t)$ as a representation of the fraction of nodes carrying a replica of $d$, by running a series of experiments where $N=2500$ nodes execute the shuffle protocol, and their caches are monitored for the presence of $d$. Unlike the experiments in Sec.~\ref{sec:expeval}, we assume full connectivity; that is, for each node, all other nodes are within reach. After 1000 rounds, where items are disseminated and replicated, a new item $d$ is inserted at a random node, at time $t=0$. We track the number of replicas of $d$ for the next 1000 rounds. The experiment is repeated 100 times and the results are averaged. These simulation results (average and standard deviation) for the protocol, together with $x(t)$, are presented in Fig.~\ref{fig:replicas}. This figure shows the same initial increase in replicas after $d$ has been inserted, and in all cases the steady state reaches precisely the expected value $\frac{c}{n}$ predicted from the stationary solution.

We repeat the calculation from Sec.~\ref{sec:optimal-s}, but now against $x(t)$, to determine which size of the exchange buffer yields the fastest convergence to the steady-state for both replication and coverage. That is, we search for the $s$ that maximizes the value of $x(t)$. We first compute the derivative of $x(t)$ with respect to $s$ ($z(t,s)$), and then derive the value of $s$ that maximizes $x(t)$, by taking \(z(\cdot,m) = \frac{\partial x}{\partial s} |_m = 0\): \(z(t,s) = \frac{\partial x}{\partial s} = \frac{2 e^{k t} (c N-n) (c n+s (-2 n+s)) t}{\left(c N+\left(-1+e^{k t}\right) n\right)^2 (n-s)^2} \text{, where }k = 2\frac{s}{c}\frac{c-s}{n-s}\). Let $z(t,s) = 0$. For $t > 0$, $cn = s(2n - s)$. Solving this equation we get $s = n \pm \sqrt{n(n-c)}$. Taking into the account that $s\leq n$, the only solution is $s = n - \sqrt{n(n-c)}$. So this coincides with the optimal exchange buffer size found in Sec.~\ref{sec:optimal-s}.

\subsection{Coverage}

We use the term coverage to denote the percentage of nodes in the network that have seen item $d$ from the moment it was introduced into the network. Let $y(t)$ represent the coverage of $d$ at time $t$. The variation in coverage per time unit, $\frac{d y}{dt}$, is determined by the fraction of nodes that have not seen $d$, $1-y$, that interacts with nodes that have $d$ in their cache, $x$. Let $\un \in \{0,1\}$, then:
\begin{eqnarray}
\frac{dy}{dt} & = & P(1\un|01) \cdot P(\bun1|\un1) \cdot (1-y) \cdot x \nonumber\\
&+&  P(1\un |01) \cdot P(1\un|1\bun) \cdot x \cdot (1-y)
\label{eq:coverage}
\end{eqnarray}

\noindent
The first term is represents increased coverage due to nodes discovering $d$ after interacting with nodes that have $d$ in their cache. This can occur when a node initiates the exchange ($P(1\un|01)$), or when the node is contacted ($P(\bun1|10)$). The second part of these terms represents the case when a node discovers and does not give away its copy of $d$ within the same round to another node. This is because coverage is only measured at the end of a gossiping round, meaning that a node that sees item $d$ for the first time, and drops it in the same round, is considered not to have seen item $d$ yet.\footnote{The reason for this is that the application has an opportunity to read from the lower-level cache only once every round.} Since nodes shuffle, on average, twice per round (once when they initiate the shuffle and again if they are contacted by a neighbour), this could occur under two scenarios: i) the node acquired $d$ by initiating an exchange with a node that had $d$ ($P(1 \un |01)$) and next lost its copy of $d$ when shuffling with a node that contacted it ($P(\bun 1|\un 1)$), or ii) the node was first contacted by a node that sent a copy of $d$ ($P(\bun 1|10)$) and next initiated a shuffle and gave away its copy of $d$ ($P(1 \un |1 \bun )$). Note that $P(\bun 1|\un 1) = 1 - P(01|10)$, as the probability $P(01|10)$ is the only transition probability that does not match the pattern $P(\bun 1|\un 1)$. Due to the symmetry of both gossiping nodes, $P(\bun 1|\un 1) = P(1 \un |1 \bun )$.

Substituting \eqref{eq:solreplica} into \eqref{eq:coverage}, we obtain
\begin{eqnarray}
\frac{dy}{dt}= 2 \cdot \frac{s}{c} \cdot \left( 1-\frac{s}{c} \cdot \frac{n-c}{n-s} \right)  \cdot (1 - y) \cdot \frac{e^{\alpha t}}{(N-\frac{n}{c})+\frac{n}{c}e^{\alpha t}} \notag
\end{eqnarray}

\begin{figure}[htb]
        \begin{center}                                            
        \begin{tabular}{c}
        \begin{minipage}[c]{\textwidth}
	\begin{center}
        \begin{tabular}[c]{c c}
        \includegraphics[scale=0.50]{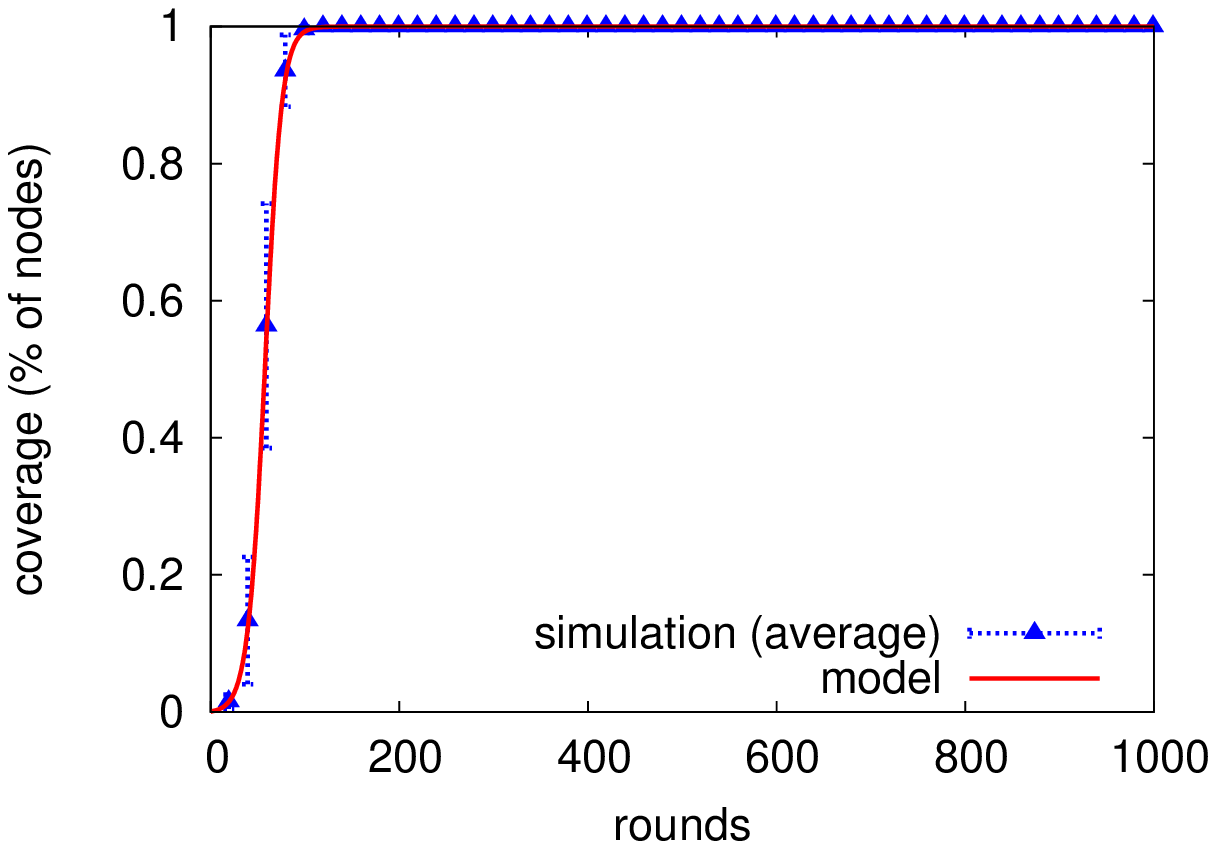}  
	& \includegraphics[scale=0.50]{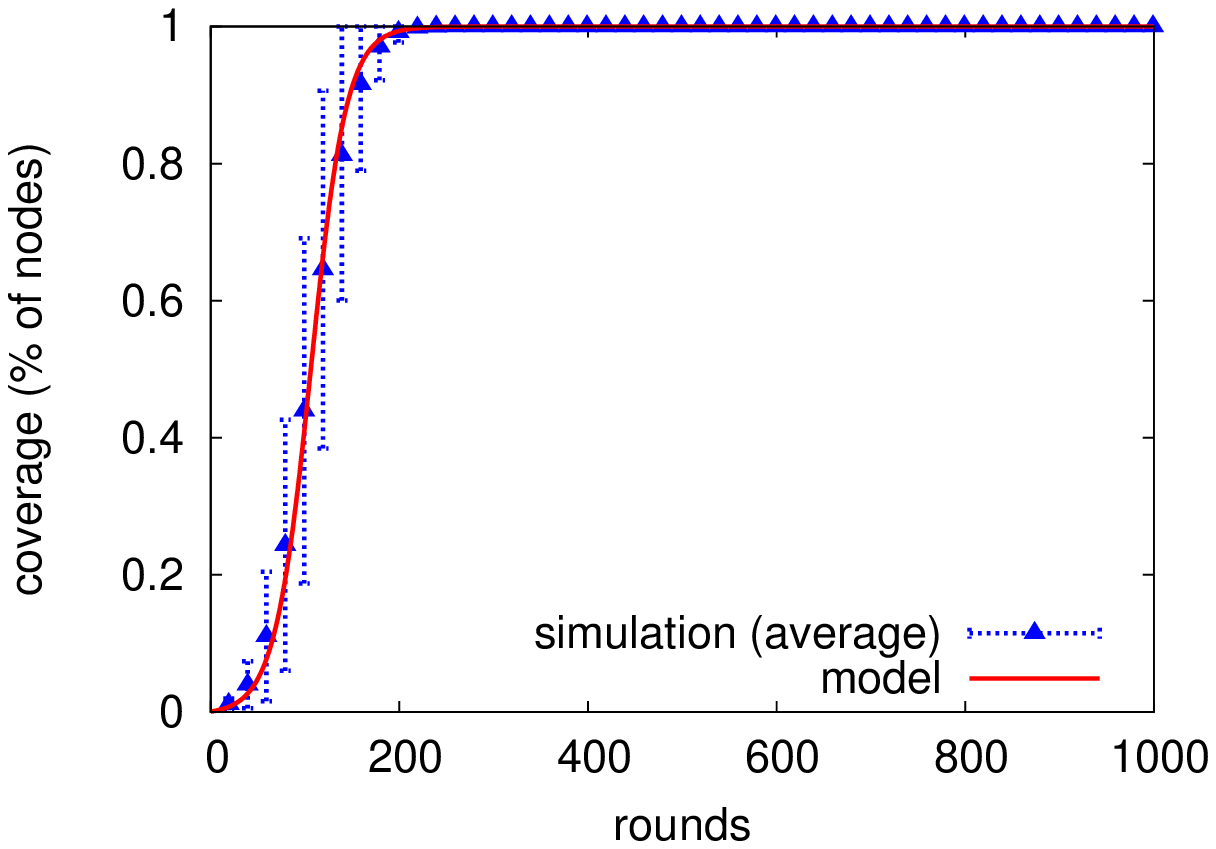} \\
	500 items & 1000 items \\
	\end{tabular}
	\end{center}
        \end{minipage} \\
        \begin{minipage}[c]{\textwidth}
	\begin{center}
        \begin{tabular}[c]{c}
        \includegraphics[scale=0.50]{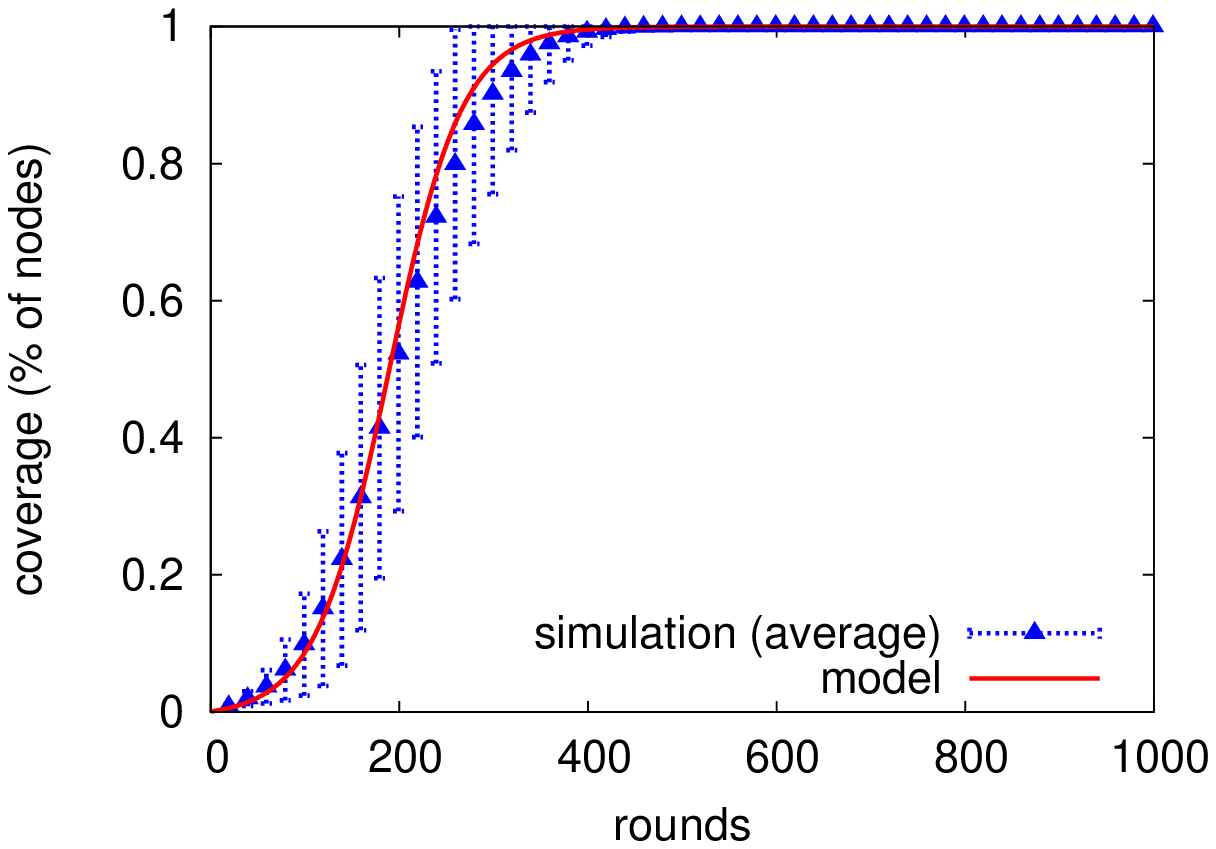} \\
	2000 items \\
	\end{tabular}
	\end{center}
        \end{minipage} \\
        \end{tabular}
	\end{center}
        \caption{Percentage of nodes in the network that have already seen a replica of item $d$, for $N=2500$, $c = 100$, $s = 50$, and $n=500$, $n=1000$ or $n=2000$.}
        \label{fig:coverage}
\end{figure}

The solution of this equation, taking into account that $y(0)=\frac{1}{N}$, is
\begin{equation*}
y(t) = 1-(N-1) N^{\beta-1} \left( \left(N - \frac{n}{c} \right) + \frac{n}{c} \cdot e^{\alpha t} \right)^{-\beta}
\end{equation*}
where $\beta$ denotes $\frac{\frac{n}{c} \cdot \left( 1-\frac{s}{c} \cdot \frac{n-c}{n-s} \right)}{\frac{c-s}{n-s}}$. 
By imposing stationarity $\frac{d y}{dt}=0$, we find the stationary solution $1$, meaning
that eventually all nodes will see $d$.

In order to evaluate how closely $y(t)$ models coverage, we use the traces from the simulations executed for Sec.~\ref{subsec:replication}. At every round, the nodes that carry a replica of $d$ are identified, and a record of the nodes that have seen $d$ since it was published is kept. Fig.~\ref{fig:coverage} presents the coverage measured for four sets of experiments, each set with a different value for $n$. As $n$ increases, a newly inserted item requires more time to cover the whole network. This is due to having more competition from other items to create replicas in the limited space available, as was previously shown in Fig.~\ref{fig:replicas}. However, as predicted by the stationary solution, in all cases the coverage eventually reaches 1. As shown in Fig.~\ref{fig:coverage}, the solution $y(t)$ models the behaviour observed in simulations, falling nicely within the standard deviation of the simulation results.

%% file: conclusion.tex
\section{Conclusions}
\label{sec:conclusions}

In this paper, we have demonstrated that it is possible to model a gossip protocol through a rigorous probabilistic analysis of the state transitions of a pair of nodes engaged in the gossip. We have shown, through an extensive simulation study, that the dissemination of a data item can be faithfully reproduced by the model. Having an accurate model of node interactions, we have been able to carry out the following:
\begin{itemize}
\item After finding precise expressions for the probabilities involved in the model, we provide a simplified version of the transition probabilities. These simplified, yet accurate, expressions can be easily computed, allowing us to simulate the dissemination of an item without the complexity of executing the actual shuffle protocol. These simulations use very little state (only some parameters and variables, as opposed to maintaining a cache) and can be executed in a fraction of the time required to run the protocol.
\item The model reveals relationships between the parameters of the system. Armed with this knowledge, we successfully optimize one of the parameters (the size of the exchange buffer) to obtain the fastest convergence of the observed properties.
\item Under the assumption of full connectivity, we are able to use the transition probabilities to model the properties of the dissemination of a generic item. Each property is ultimately expressed as a formula which is shown to display the same behaviour as the average behaviour of the protocol, verifying the validity of the model.
\end{itemize}
While gossip protocols are easy to understand, even for a simple push/pull protocol, the interactions between nodes are unexpectedly complex. Understanding these interactions provides insight into the mechanics behind the emergent behaviour observed in gossip protocols. We believe that understanding the mechanics of gossiping is the key to optimizing (and even shaping) the emergent properties that make gossiping appealing as communication paradigm for distributed systems.